\documentclass[a4paper,twoside]{article} % default 10pt

\input{vssc.sty} 

\newcommand\V{\emph{V}}              
              
\newcommand\TG{\emph{TG}}              

\begin{document}

%%%%%%%%%%%%%%%%%%%%%%%%%%%%%%%%%%%%%%%%%%%%%%%%%%%%%%%%%%%%%%%%%%%%%%%%%%%%%%
%%
%% Title here
\OBStitle{Betelgeuse -- A Century and more of Variation}

%%%%%%%%%%%%%%%%%%%%%%%%%%%%%%%%%%%%%%%%%%%%%%%%%%%%%%%%%%%%%%%%%%%%%%%%%%%%%%
%%
%% Author list here
\OBSauth{
	Christopher Lloyd
}
	
%\OBSauth{
%	David Crosby\,$^{1,2,3}$,
%	Stephen Stills\,$^{2,4}$,
%	Graham Nash\,$^{2,5}$, and \\
%	Neil Young\,$^{2,6,7}$
%	}
	
%%%%%%%%%%%%%%%%%%%%%%%%%%%%%%%%%%%%%%%%%%%%%%%%%%%%%%%%%%%%%%%%%%%%%%%%%%%%%%
%%
%% Institution/affiliation here

%	\OBSinstone{School of Mathematical and Physical Sciences, University of Sussex}

%	\OBSinst{Experimental Optics, University of Isengard,  Nan Curun{\'i}r}
%	\OBSinst{School of Mathematical and Physical Sciences, University of Sussex}

%%%%%%%%%%%%%%%%%%%%%%%%%%%%%%%%%%%%%%%%%%%%%%%%%%%%%%%%%%%%%%%%%%%%%%%%%%%%%%
%%
%% Abstract here
\OBSabstract{
The mean light curve of Betelgeuse is constructed from the visual data in BAA VSS and AAVSO archives. Period analysis reveals clusters of periods around 2000 and 400 days but these are swamped by the long-term trends. No identifiable periods emerge but the feature near 400 days is the most persistent and survives even when the range of variation is low. Herschel's data from 1836-40 and early data from the BAA VSS around 1900 show a range of V {\raise.3ex\hbox{\boldmath${\scriptstyle\sim}$}}  0 -- 1, so the star was brighter and more active than recently. Historically it shows a wide range of behaviour.
}

%\section*{Introduction}

On a good day Betelgeuse is one of the brightest stars in the night sky, brighter that Procyon, Rigel and Capella, even Vega and possibly even brighter than Arcturus, leaving only Rigel Kent, Canopus and Sirius outshining it. But these days are rare, and there are also bad days as witnessed recently when Betelgeuse underwent an unusually deep and rapid fade down to the relegation zone of magnitude 1.6. Betelgeuse spends most of its time at magnitude 0.5-0.8 in the realm between Procyon and Aldebaran, but it does undergo slow excursions of half a magnitude with brief runs up to magnitude zero and occasional rapid fades. Its spectral type is M1-2\,1a-ab so it is a very luminous early M-type supergiant, which accounts for its pronounced orange colour as opposed to the deeper red of most Mira-type variables. Betelgeuse itself is classified as an SRc variable due to its small range of variation and lack of clear periodicity. The AAVSO 
\href{https://www.aavso.org/vsx/index.php?view=detail.top&oid=24710}{VSX}
gives a main period of 423 days, with a secondary at $\sim 2100$ days, but the 
\href{http://www.sai.msu.su/gcvs/cgi-bin/search.cgi?search=alf+Ori}{GCVS} gives the period as 2335 days. A trawl through the literature will turn up many others.

Betelgeuse is such a bright star in a prominent constellation that its variability must have been known to the ancients. Wilk \cite{1999JAVSO..27..171W}
continues the theme that the variability of bright stars, including Betelgeuse, was known to the pre-Classical Greeks and is reflected in their mythology. Myths of the Australian Aboriginals lead to an even more direct description of its variability with the star becoming periodically brighter and fainter (see 
%\href[http://articles.adsabs.harvard.edu/pdf/2014JAHH...17..180L}{
Leaman \& Hamacher \cite{2014JAHH...17..180L}).

When it comes to more modern times it seems most likely that the variation was discovered or rediscovered, in the western world at least, by Sir John Herschel as reported to the RAS in 1840. Herschel started making naked-eye observations of some dozens of stars at the Cape, including Betelgeuse during 1836, and continued these the following years, including on the ship during the trip back to the UK in 1838, until early 1840. The results are presented in a 
\emph{Monthly Notices} paper \cite{1840MNRAS...5...11H}
%https://ui.adsabs.harvard.edu/abs/1840MNRAS...5...11H/abstract
and in much more detail in the 
\emph{Memoirs} \cite{1840MmRAS..11..269H}. 
%https://ui.adsabs.harvard.edu/abs/1840MmRAS..11..269H/abstract
Herschel was following his own advice in making observations of variable stars, which he thought showed 
\emph{``a sure promise of rich discovery"}, and this is as true today as it was when he wrote his 
\emph{Treatise on Astronomy}
\cite{1833tras.book.....H} in 1833. 
%https://ui.adsabs.harvard.edu/abs/1833tras.book.....H/abstract

\begin{table}[t]
	\caption{Herschel's observations of Betelgeuse from 1836 to 1840. The bracketing stars only are given for the early observations but the \emph{``imaginary stars"} indicated ``\textbar" are included for the later data. The derived magnitudes are based on modern \V\ measurements and additional notes to the observations. There are some inconsistencies due to extinction effects and some small rounding issues.\label{tab:herschel}} % no space
	\centering
	\begin{tabular}{cclc}
		\hline
		\addlinespace[2pt]
			Date			&JD			&Measure									&Mag. \\
		1836 Mar. 22	&2391726	&$\alpha$ Crucis -- Betelgeuse -- Regulus		&1.1 \\
		1836 Nov. 12	&2391961	&Betelgeuse\} -- Procyon &\\
		&			&Rigel\}													&0.0 \\
		1836 Nov. 13	&2391962	&Betelgeuse = Rigel							&0.1 \\
		1836 Nov. 26	&2391975	&Procyon -- Betelgeuse -- Achernar			&0.4 \\
		1837 Oct. 24	&2392307	&Betelgeuse -- Achernar (high in the sky) 	&0.3 \\
		&2392307	&Betelgeuse -- Rigel -- Aldebaran (low)		&--0.1 \\
		1837 Dec. 3	&2392347	&Rigel very much larger than Betelgeuse		&(0.8) \\
		1837 Dec. 16	&2392360	&Rigel -- Achernar -- Betelgeuse 				&0.7 \\
		1837 Dec. 29	&2392373	&Procyon -- Betelgeuse -- Aldebaran			&0.6 \\
		1838 Jan. 2	&2392377	&Achernar -- Betelgeuse -- Pollux				&0.8 \\
		1838 Jan. 6	&2392381	&Achernar -- Betelgeuse -- Aldebaran			&0.7 \\
		1838 Jan. 13	&2392388	&Achernar -- Betelgeuse -- Aldebaran			&0.8 \\
		1838 Feb. 25	&2392431	&Procyon -- Betelgeuse -- $\alpha$ Crucis			&0.6 \\
		1838 Apr. 14	&2392479	&Rigel -- Betelgeuse -- Aldebaran				&0.5 \\
		1839 Jan. 16	&2392756	&Aldebaran is greater than Betelgeuse		&(1.1) \\
		1839 Jan. 17	&2392757	&Aldebaran -- Betelgeuse -- Pollux			&1.0 \\
		1839 Jan. 22	&2392762	&Aldebaran -- Betelgeuse -- Pollux			&1.0 \\
		1839 Nov. 26	&2393070	&Capella \textbar\textbar\ Betelgeuse \textbar\ Rigel				&0.1 \\
		1839 Nov. 30	&2393074	&Rigel \textbar\ Procyon, Betelgeuse  \textbar\textbar\ Aldebaran	&0.4 \\
		1839 Dec. 11	&2393085	&Rigel  \textbar\ Procyon  \textbar\ Betelgeuse  \textbar\textbar\ Aldebaran	&0.6 \\
		1839 Dec. 29	&2393103	&Rigel, Procyon, Betelgeuse, Aldebaran		&0.7 \\
		1840 Jan. 2	&2393107	&Rigel \textbar\textbar\ Procyon \textbar\ Betelgeuse \textbar\textbar\ Aldebaran	&0.6 \\
		1840 Jan. 5	&2393110	&Rigel \textbar\textbar\ Procyon \textbar\ Betelgeuse \textbar\textbar\ Aldebaran	&0.6 \\
		1840 Jan. 6	&2393111	&Rigel \textbar\textbar\ Procyon \textbar\textbar\ Betelgeuse \textbar\textbar\ Aldebaran&	0.7 \\
		1840 Jan. 7	&2393112	&Procyon \textbar\ Betelgeuse  \textbar\ Aldebaran			&0.7 \\
		1840 Feb. 25	&2393161	&Procyon \textbar\textbar\ Betelgeuse \textbar\ Aldebaran			&0.7 \\
		1840 Apr. 18	&2393214	&Procyon \textbar\ Betelgeuse  \textbar\textbar\ Aldebaran			&0.6 \\
		\hline
	\end{tabular}
	%
	%\vspace{6pt}
	%a) iTelescope T24 MPC U69 Sierra Remote Observatory, Auberry, CA\\
	%b) iTelescope T11 MPC H06 New Mexico Skies, Mayhill, NM
	%
	%For notes below the table a blank line and \vspace should be used
\end{table}

\begin{figure}[t]
	\centering
	\includegraphics[width=0.7\textwidth]{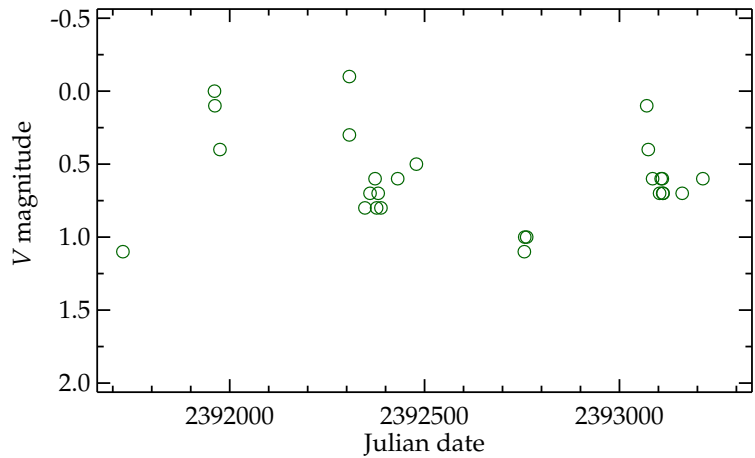}
	\caption{Herschel's light curve of Betelgeuse from 1836 to 1840. The seasonal variation is clear and the star was generally bright, often comparable to or brighter than Rigel.}
	\label{fig:herschellc}
\end{figure}

The discovery can probably be pinpointed to the 26th November 1839 when he was \emph{``surprised, and I may almost say startled by the extraordinary splendour of $\alpha$~Orionis"}, which at that time he placed between Capella and Rigel. Both have $V \sim 0.1$. He was then prompted to review his observations of the previous years and came to the realization that Betelgeuse was variable with a likely period of about one year. Herschel's method of observation was at the same time obvious and original, as no one else seemed to be doing this. In the early observations he simply recorded which two stars bracketed the variable and recorded this in the undeniably magnificent, \emph{``Order of Lustre"}, so there would be a sequence like, $\alpha$~Crucis -- Betelgeuse -- Regulus, and no magnitudes were involved. As time progressed he inserted \emph{``imaginary stars"} at equal intervals of lustre between what could be called the comparisons, and so the fractional method was born. Herschel's measurements are given in Table~\ref{tab:herschel} and plotted in Figure~\ref{fig:herschellc}. His estimates are given with the imaginary stars as he indicated ``\textbar" and the magnitudes calculated using modern values of the \V\ magnitudes, with additional guidance from the notes that accompanied each estimate. The general order of lustre that he used is, Capella (0.1), Rigel (0.1), Procyon (0.4), Achernar (0.5), Aldebaran (0.9), $\alpha$~Crucis (0.8), Pollux (1.2) and Regulus (1.4), but it is clear that some small inconsistencies exist. However, given the wide range in colour of these stars and their dispersal over the sky this is not unexpected, and has no material effect on the results. Some of the apparently inconsistent results in Table~\ref{tab:herschel} can be explained by extinction effects on either the variable or the primary comparisons.

Herschel's short run of observations span the range $V = 0$ -- 1.1, which is much larger than its normal range and reaches the bright limit, so during this time the star was bright and active. From Figure~\ref{fig:herschellc} it can be seen that if the uncertainties are small, and if you're Herschel they probably are, then a vaguely seasonal, periodic variation could fit the data, and this is what Herschel reported. The only other observation from around this time is a report by
R.H. Allen \cite{1899sntm.book.....A} % https://ui.adsabs.harvard.edu/abs/1899sntm.book.....A/abstract
and cited by Wilk \cite{1999JAVSO..27..171W}
that on 5th December 1852, Betelgeuse was reported to be \emph{``the largest (brightest) star in the northern hemisphere".}

Despite Herschel's efforts to kick-start variable star astronomy both through his manifesto in 1833 and his practical efforts in 1840 and after, there was little response from the community. Argelander started doing similar work in Bonn in the 1840s and the modern basis of the magnitude scale was established by Pogson by 1856 
(see \eg\ Jones, \cite{1968ASPL...10..145J} and %https://ui.adsabs.harvard.edu/abs/1968ASPL...10..145J/abstract
Reddy \etal \cite{2007JBAA..117..237R}). 
Over the next 30 or so years various variable star organisations came and went in the UK as thoroughly detailed by 
Toone \cite{2010JBAA..120..135T} %(2010, J.BAA 120, 135)  https://ui.adsabs.harvard.edu/abs/2010JBAA..120..135T/abstract
before the BAA VSS finally emerged in 1890. 

The earliest VSS observations of Betelgeuse were made by Jo\~{a}o De Moraes Pereira, 
\cite{2010JBAA..120..101B} one of only 19 members of the VSS at this time, who made a series of observations in the winter of 1893/4 and the following two years. Unfortunately, neither the comparison stars nor the step values he used were recorded, but the die was cast and a few years later E.E. Markwick,
\cite{2012JBAA..122..335S}
and several other early pioneers of variable star astronomy were making observations. For these also no sequences are available, but the step values have been recorded so it is possible to see the obvious comparison stars Procyon and Aldebaran being adopted, and with the same magnitudes as used today. 

\begin{figure}[p]
	\centering
	\includegraphics[width=1.\textwidth]{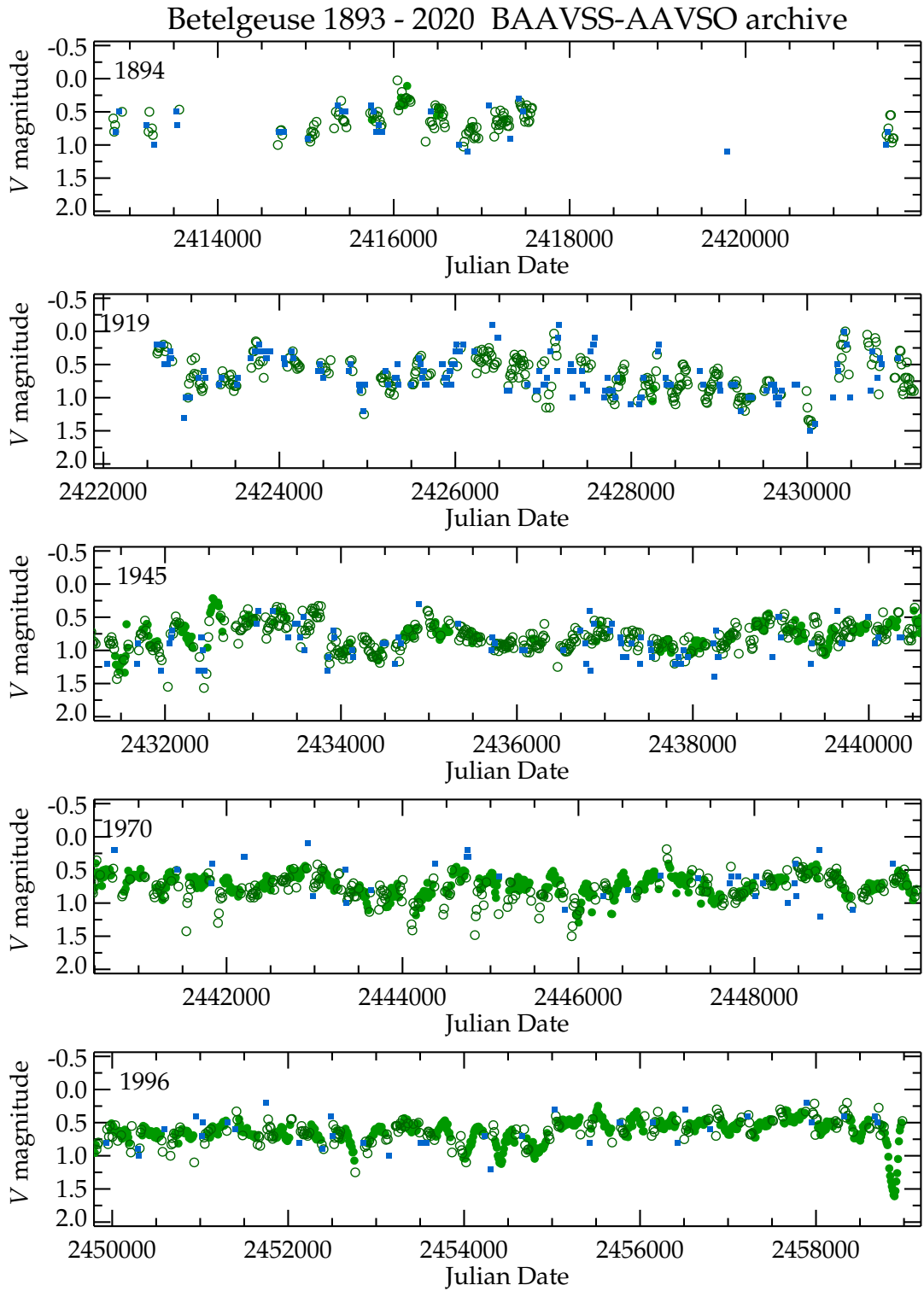}
	\caption{The combined BAA VSS -- AAVSO light curve of Betelgeuse from 1893 to the present. The points are 10-day bins with single points shown as dots, up to 10 points as open circles, and more than 10 as filled circles. The year is shown at the start of each panel which cover $\sim25$ years.}
	\label{fig:betelgeuselc}
\end{figure}

The complete light curve from the combined BAA VSS and AAVSO archives is shown in Figure~\ref{fig:betelgeuselc}. The plot shows 10-day means with a single observation as a small dot, up to 10 observations as an open circle and above 10 as a filled circle. Care must be taken when interpreting specific features as the alignment of the 10-day bins can make a large difference to the number of observations that are included in each bin. Not surprisingly the early data are relatively sparse and there is a large gap from 1910 to 1920. Following that the record is a largely complete, but it is thin in places. Even from the start the seasonal trends noted by Herschel can be seen and the range of variation from $V \sim 0 - 1$ is also similar to what he reported. There are also signs of longer time-scale features with the star cycling over $\sim$ 2000 days. From JD $\sim$ 2427000 the star begins to have significant excursions below magnitude 1, with a few deeper fades to $V \sim 1.5$, and consistent brightenings to magnitude 0. Over this period the seasonal trends become a clearer, saw-tooth pattern as more observations are made. Around JD = 2431000 the star undergoes a substantial fade from $V \sim 0.3 - 1.4$ and this was followed $\sim$ 500 and 1000 days later by two other deep fades which coincide with the minima of the saw-tooth pattern. These are the most consistent and faintest magnitudes in the record -- up until the present -- but they are not particularly well observed. There is then a long section up to JD $\sim$ 2450000 where the saw-tooth pattern becomes less visible and the light curve is largely dominated by small variations on long time scales. For the last 25 years the range of variation has been at a minimum and a weak saw-tooth pattern has re-emerged. Also, there has been a distinct change in level at JD $\sim$ 2455000 with mean $V = 0.67$ before and $V = 0.56$ after, despite the recent deep fade. Mean \V\ magnitude for 4000-day sections of the 10-day mean light curve are given in Table~\ref{tab:means}.

\begin{figure}[t]
	\centering
	\includegraphics[width=1.\textwidth]{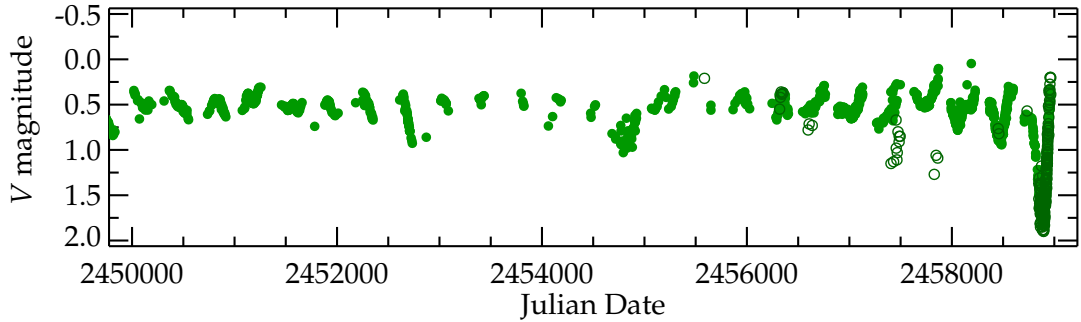}
	\caption{The recent \V\ and \TG\ data of Betelgeuse covering the same period as the bottom panel of Figure~\ref{fig:betelgeuselc}.}
	\label{fig:betelgeusevlc}
\end{figure}

That is why Betelgeuse is an SRc variable. There is no clear periodicity but some features of the light curve do emerge, 1) the long-term changes in level, 2) the $\sim$2000+ day variation which comes and goes, and 3) the saw-tooth pattern which is sometimes equally invisible. In recent years an increasing number of photoelectric and CCD \V-band measurements have been made together with some DSLR \TG\ values. These data from JD $\sim$ 2450000 are shown in Figure~\ref{fig:betelgeusevlc} and cover the same range as the bottom panel of Figure~\ref{fig:betelgeuselc}. With less scatter the seasonal runs are much more clearly visible and surprisingly small details in the visual record can be confirmed. Generally, the comparison is remarkably good. Curiously though, the small change of level in the visual data cannot be seen, but the latter half of the \V\ data do have a slightly brighter limit. The other important point to emerge from the \V\ data is that the variations are almost entirely linear, which is saying something about the processes involved.

\begin{table}[!b]
	\caption{Mean magnitudes and dispersions for 4000-day intervals of the 10-day means.\label{tab:means}} % no space
	\centering
	\begin{tabular}{ccc}
		\hline
		\addlinespace[2pt]
		Julian Date range	&Mean&	Standard  \\
		&	& deviation \\
		2412000 -- 2416000	&0.67&	0.16 \\
		2416000 -- 2420000	&0.58&	0.22 \\
		2420000 -- 2424000	&0.59&	0.26 \\
		2424000 -- 2428000	&0.59&	0.24 \\
		2428000 -- 2432000	&0.82&	0.29 \\
		2432000 -- 2436000	&0.75&	0.23 \\
		2436000 -- 2440000	&0.82&	0.17 \\
		2440000 -- 2444000	&0.71&	0.17 \\
		2444000 -- 2448000	&0.78&	0.20 \\
		2448000 -- 2452000	&0.67&	0.13 \\
		2452000 -- 2456000	&0.67&	0.17 \\
		2456000 -- 2460000	&0.56&	0.21 \\
		\hline
	\end{tabular}
	%&&
	%\vspace{6pt}&&
	%a) iTelescope T24& MPC& U69 Sierra Remote Observatory, Auberry, CA\\
	%b) iTelescope T11& MPC& H06 New Mexico Skies, Mayhill, NM
	%
	%For notes below the table a blank line and \vspace should be used
\end{table}

\begin{figure}[t]
	\centering
	\includegraphics[width=0.7\textwidth]{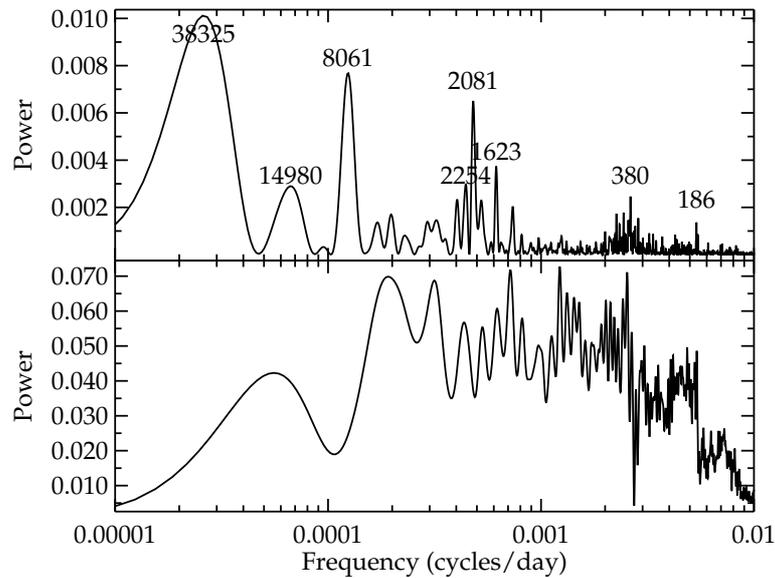}
	\caption{(Top) The DFT periodogram of the 10-day means with the dominant periods identified and (Bottom) the DFT of the \V\ data.}
	\label{fig:dft}
\end{figure}

There are various tools that can be used to find periodic variations, even if they are buried in the noise, but the interpretation is often the problem. The most widely used technique is probably the Discrete Fourier Transform (DFT) periodogram which measures the power -- that is the semi-amplitude2 -- at a particular frequency. The DFT of the 10-day means is shown in the upper panel of Figure~\ref{fig:dft} with the periods associated with the major peaks identified. The first point to notice is the power range. The peak corresponds to an amplitude of $\pm0.1$ magnitudes so no periods here are going to account anything but the smallest variation in the light curve. The longest period is longer than the run of data and is driven by the small changes of level identified in Table~\ref{tab:means}, and the two other long periods are a response to other long-term trends, which are invisible to the eye. The first period that can be clearly related to variations in the light curve is at 2081 days, but this is close to two other prominent periods and a cluster of weaker ones between 1000 and 3000 days. All of these are related to the continuum of variations on this time scale in the data, but the specific periods that appear dependent on the vagaries of that particular data set. Similarly for the features near 380 days. These correspond to the saw-tooth pattern and cover a range of 300 -- 500 days. While this is not periodic it is probably the most persistent and diagnostic feature of the light curve. The shortest period identified is a harmonic of this one.

The DFT of the \V\ data is shown in the lower panel of Figure~\ref{fig:dft} and immediate differences can be seen. The power is much larger, and accounts for most of the variation seen in the data, and the only really persistent feature is the one near 400 days. The 2000+ day cluster has gone because the recent light curve has been essentially flat, and there is little power at the longest periods because there are no data on that time scale.

There is clearly no single period near 2000 or 400 days but both could be representative of real periods in the data that meander between 1000 -- 3000 and 300 -- 500 days. The cluster of features in the periodogram just gives a snapshot of this for the past 100 years, that would eventually become filled in over great tracts of time. The alternative is that there is no periodicity at all and that the processes at work in the star recognize no clock, just a loose time scale -- sometimes. The difference in interpretation leads to very different visions of how the atmosphere of the star operates.

The recent deep fade has led to a renewed interest in Betelgeuse and an explosion in the number of observations made, if not the star itself (see Figure 2 of the 
\href{https:www.britastro.org/vss/VSSC184.pdf#page=14}{accompanying paper}\ \cite{2020VSSCir.184..14K}). The fade was first noticed by 
Guinan \etal, \cite{2019ATel13341....1G} \href{http://www.astronomerstelegram.org/?read=13341}{(ATel\#13341)}  
who also pointed out that it appeared to be a deep minimum of the 400 day cycle. In a later update 
\href{http://www.astronomerstelegram.org/?read=13365}{(ATel\#13365)} \cite{2019ATel13365....1G}
they also suggested that the fade might be due to the coincidence of the minima from both the 400 and 2000+ day cycles, not in an additive sense as the amplitudes are too low, but presumably as some sort of resonance. However, even this seems unlikely as the ephemeris of the long cycle must be unknown; they put an uncertainty of 180 days on the period, and it has been effectively absent for the past 25 years.

Obviously, there is great interest in what is likely to happen next year. The recent minimum was in early February so the next one might reasonably be in March 2021; periodic or not there is still a preferred time scale. Betelgeuse is on average as bright as it's ever been and in the past this has usually been accompanied by greater activity. Maybe unprecedented bright magnitudes will be seen. Looking back at the previous double deep minima around 1945 raises the tantalizing prospect of further deep minima disappearing into the twilight. Were they real -- will it do it again?

In view of recent events this comment by Herschel is equally apt today.

\emph{``It may be easily supposed ... that the confirmation or disappointment of this expectation is awaited with no small interest."}

Indeed. All eyes will be on Betelgeuse next year.

%% Add references as required:
%\bibliographystyle{vssc}
%\bibliography{betelgeuse}

\begin{thebibliography}{10}
	\newcommand{\enquote}[1]{`#1'}
	
	\bibitem{1999JAVSO..27..171W}
	S.~R. {Wilk},
	\textit{\href{https://ui.adsabs.harvard.edu/abs/1999JAVSO..27..171W}{\jaavso}},
	\textbf{27}, 171, 1999
	
	\bibitem{2014JAHH...17..180L}
	T.~M. {Leaman} \& D.~W. {Hamacher},
	\textit{\href{https://ui.adsabs.harvard.edu/abs/2014JAHH...17..180L}{Journal
			of Astronomical History and Heritage}}, \textbf{17}, 180, 2014
	
	\bibitem{1840MNRAS...5...11H}
	J.~F.~W. {Herschel},
	\textit{\href{https://ui.adsabs.harvard.edu/abs/1840MNRAS...5...11H}{\mnras}},
	\textbf{5}, 11, 1840
	
	\bibitem{1840MmRAS..11..269H}
	J.~F.~W. {Herschel},
	\textit{\href{https://ui.adsabs.harvard.edu/abs/1840MmRAS..11..269H}{\memras}},
	\textbf{11}, 269, 1840
	
	\bibitem{1833tras.book.....H}
	J.~F.~W. {Herschel},
	\textit{{\href{https://ui.adsabs.harvard.edu/abs/1833tras.book.....H}{A
				Treatise on Astronomy}}} (Longman, Rees, Orme, Brown, Green, \& Longman, and
	John Taylor), 1833
	
	\bibitem{1899sntm.book.....A}
	R.~H. {Allen},
	\textit{{\href{https://ui.adsabs.harvard.edu/abs/1899sntm.book.....A}{Star-names
				and their meanings}}} (G.E. Stechert), 1899
	
	\bibitem{1968ASPL...10..145J}
	D.~{Jones},
	\textit{\href{https://ui.adsabs.harvard.edu/abs/1968ASPL...10..145J}{Leaflet
			of the Astronomical Society of the Pacific}}, \textbf{10}, 145, 1968
	
	\bibitem{2007JBAA..117..237R}
	V.~{Reddy}, K.~{Snedegar} \& R.~K. {Balasubramanian}, \textit{\jbaa},
	\textbf{117}, 237, 2007
	
	\bibitem{2010JBAA..120..135T}
	J.~{Toone},
	\textit{\href{https://ui.adsabs.harvard.edu/abs/2010JBAA..120..135T}{\jbaa}},
	\textbf{120}, 135, 2010
	
	\bibitem{2010JBAA..120..101B}
	V.~{Bonif{\'a}cio}, I.~{Malaquias} \& J.~{Fernandes},
	\textit{\href{https://ui.adsabs.harvard.edu/abs/2010JBAA..120..101B}{\jbaa}},
	\textbf{120}, 101, 2010
	
	\bibitem{2012JBAA..122..335S}
	J.~{Shears},
	\textit{\href{https://ui.adsabs.harvard.edu/abs/2012JBAA..122..335S}{\jbaa}},
	\textbf{122}, 335, 2012
	
	\bibitem{2020VSSCir.184..14K}
	M.~{Kidger},
	\textit{\href{https://www.britastro.org/vss/VSSC184.pdf\#page=14}{\vssc}},
	\textbf{184}, 14, 2020
	
	\bibitem{2019ATel13341....1G}
	E.~F. {Guinan}, R.~J. {Wasatonic} \& T.~J. {Calderwood},
	\textit{\href{https://ui.adsabs.harvard.edu/abs/2019ATel13341....1G}{The
			Astronomer's Telegram}}, \textbf{13341}, 2019
	
	\bibitem{2019ATel13365....1G}
	E.~F. {Guinan}, R.~J. {Wasatonic} \& T.~J. {Calderwood},
	\textit{\href{https://ui.adsabs.harvard.edu/abs/2019ATel13365....1G}{The
			Astronomer's Telegram}}, \textbf{13365}, 2019
	
\end{thebibliography}

%%%%%%%%%%%%%%%%%%%%%%%%%%%%%%%%%%%%%%%%%%%%%%%%%%%%%%%%%%%%%%%%%%%%%%%%%%%%%%
%%
%% VSSC reference here
\VSSCref{184, (2020)}

\end{document}